\documentclass[aps,prl,showpacs,twocolumn,superscriptaddress]{revtex4}
\usepackage{graphicx}
\usepackage{amsmath}

\begin{document}

\title{Theoretical Study of Magnetic Moments Induced by Defects at the SiC (110) Surface}
\author{Adrien Poissier}
\affiliation{
Physics and Astronomy, State University of New York, Stony Brook, New York 11794-3800, USA}
\author{Nicol\'as Lorente}
\affiliation{
 Centre d'Investigaci\'o en Nanoci\`encia  i
Nanotecnologia (CSIC-ICN), Campus de la UAB, E-08193 Bellaterra, Spain}
\author{F\'elix Yndurain}
\affiliation{Departamento de F\'{\i}sica de la Materia Condensada,
Universidad Aut\'onoma de Madrid, 28049 Madrid, Spain}
\date{\today}
\begin{abstract}
The effect of different surface defects on the atomic and electronic structures of cubic $\beta$-SiC(110) surface are studied by means of a first principles calculation based on Density Functional Theory using the SIESTA code. In the calculations, the possibility of different spin population at each atom is allowed. We find that while adsorption of oxygen or nitrogen and adsorption of hydrogen at the C surface carbon atoms do not induce magnetic moments on SiC(110); Si vacancies, substitutional C at the Si site and H or F adsorbed at the silicon surface sites induce localized magnetic moments as large as 0.7 $\mu_{B}$ at the carbon atoms close to the defect. The local magnetic moment arrangement varies from ferromagnetic in the case of H adsorption to antiferromagnetic in the Si vacancy and substitutional C cases. The case of H adsorption at the Si surface atoms is discussed in detail. It is concluded that magnetism is mainly due to the local character of the C valence orbitals. 
\\

\end{abstract}

\pacs{71.20.Mq,75.70.Rf, 81.05.Uw }
\maketitle

The localized character of the carbon valence charge has motivated the study of the possibility of carbon atoms developing localized magnetic moments near defects \cite{Magnetic-Carbon}. This has prompted the study of a variety of defects at different carbon atomic configurations like graphite \cite{Chema},  graphene \cite {Graphene-Guinea,Choi}, nanotubes \cite {Yuchen-Ma, Nieminen} and other configurations \cite{Hohne-Esquinazi,Esquinazi-Hohne} including defect-free diamond surfaces \cite {C-(110)}. 
Here, in this work, we consider the appearance of localized magnetic moments near defects on the (110) surface of cubic silicon carbide ($\beta$-SiC). The quasi one dimensional character of the Si-C chains at the surface and the localized character of the carbon atomic valence charge suggest the possibility of defect induced magnetism. 

Silicon carbide has been considered for many years as a compound with important potential practical applications like power electronics, heterogeneous catalysis support, structural and protective components for use in future nuclear fusion reactors, etc. This interest has been fostered in the last years by the possibility to obtain SiC in different atomic configurations ranging from macro molecules like fullerenes to two-dimensional sheets and their wrapped configuration in nanotubes (see for instance the work of Melinon {\em et al.}~\cite{Melinon-Nature} and references therein). 

\begin{figure} [ht]
\includegraphics[width=\columnwidth]
{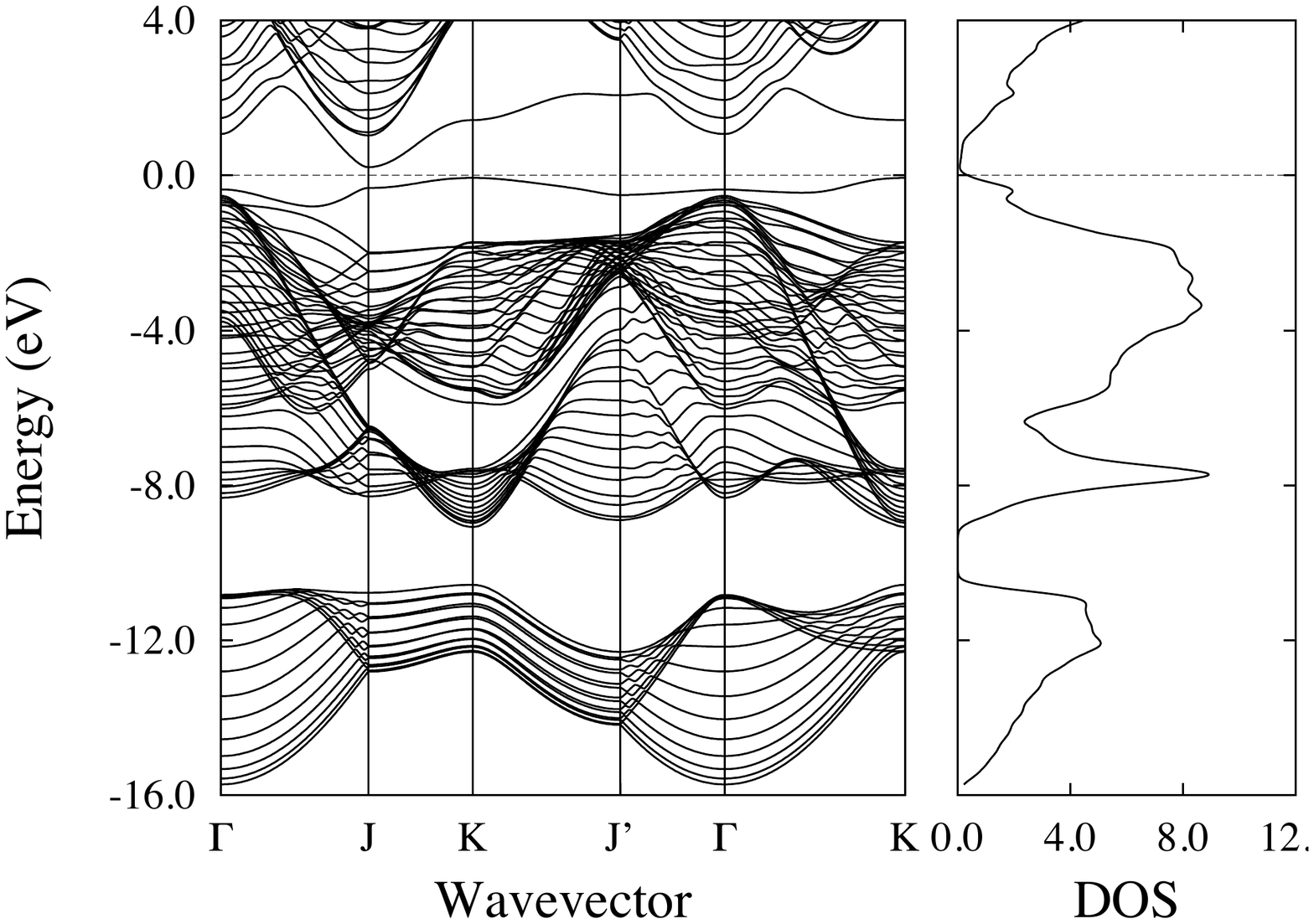}
\caption{Calculated band structure an total density of states for a 15-layers slab of  SiC(110) with a buckled, unreconstructed clean surface. The surface states band structure is apparent at the bulk band gap. The origin of energies is taken at the Fermi Energy.}
\label{bands}
\end{figure} 

The free SiC (110) surface has been already theoretically studied by Sabisch et al. \cite{Pollmann} (for a complete review of the different SiC surfaces see reference \cite{Pollmann-Review}). Unlike other surface orientations, the free SiC(110) surface does not present any reconstruction as different theoretical works reveal \cite{Pollmann-Review}. To the best of our knowledge there are no published experimental studies of this surface. 

\begin{figure} 
\includegraphics[width=\columnwidth]
{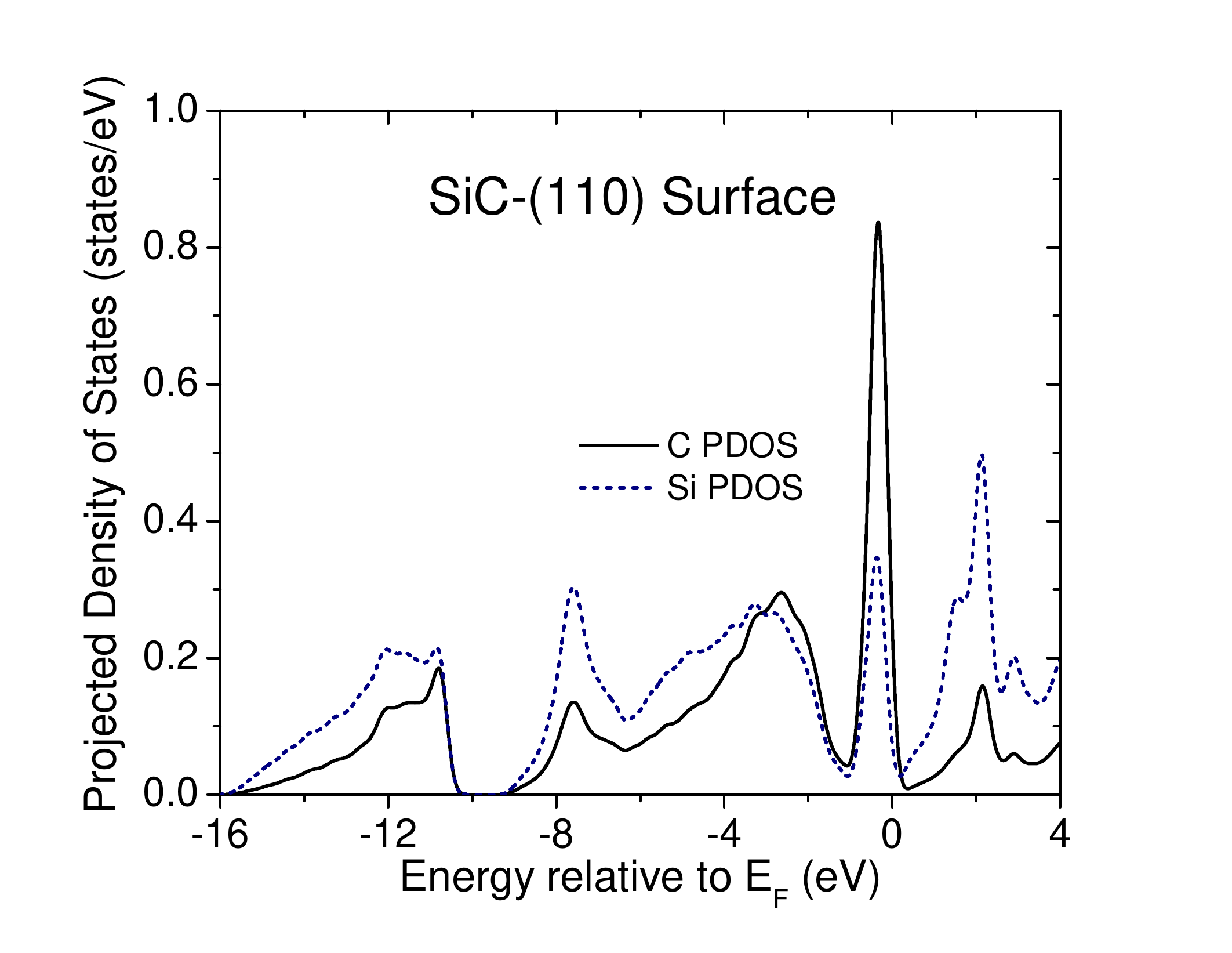}
\caption{ (Color online) Projected electronic densities of states at the surface C (solid black line) and Si (broken blue line) atoms in the SiC(110) surface. The origin of energies is taken at the Fermi Energy. A small imaginary contribution to the energy has been added for presentation purposes.}
\label{Surface-Densities}
\end{figure} 
Here, in this work, we present the results of a first principles calculation of different defects at the SiC(110) surface. The calculations throughout this work were performed within the density functional  theory (DFT)\cite{kohn}, using the generalized gradient approximation (GGA) \cite{pbe}  for the exchange and correlation. The calculations were obtained with the SIESTA~\cite{Siesta1, Siesta2}  method, which uses a basis of numerical atomic orbitals~\cite{san} and separable~\cite{kle} norm conserving pseudopotentials~\cite{tro} with partial core corrections~\cite{lou}. We have found satisfactory the standard double-$\zeta$ basis with polarization orbitals (DZP) which has been used throughout this work. The bulk calculation yields a lattice constant of 4.41 \rm{\AA} (Si-C bond length of 1.91 \rm{\AA}) in fair agreement with the experimental value of 4.36 \rm{\AA}. We find a partially ionic Si-C bond with a charge transfer of 0.45 electrons to the Si atom as evaluated using Mulliken population
analysis. We obtain a band gap of 1.31 eV similar the 1.29 eV calculated by  Sabisch et al. \cite{Pollmann} and to be compared with the experimental one of 2.417 eV \cite{gap}.

To simulate the free SiC(110) surface we have performed calculations of different size slabs formed by the stacking of (110) planes. In particular, we have considered in this work slabs built by 7 and 15 layers. In all cases the ``back´´ surface was properly saturated with hydrogen atoms. The convergence of the relevant precision parameters was carefully checked. The real space integration grid had a cut-off of 400 Ryd. Of the order of up to 600 $k$ points were used in the two-dimensional Brillouin zone sampling using the Monkhorst-Pack k-points sampling. 
To accelerate the self-consistency convergence, a polynomial broadening of the energy levels was performed using the method of Methfessel and Paxton \cite{met} which is very suitable for systems with a large variation of the density of states at the vicinity of the Fermi level which is the case in our system (see below). Broadening like Fermi-Dirac can be inappropriate and give wrong results.  It is worth to mention that the energy differences between non-magnetic and magnetic solutions are, in general, small, what requires a very high convergence in all precision parameters and tolerances. To obtain the equilibrium geometry we relaxed the atoms until the forces acting on them were smaller than 0.01 eV/\rm{\AA}.
 
Results of the calculations of a 15 layers slab are reported in Figures \ref{bands} and  \ref{Surface-Densities}. We, like in reference \cite{Pollmann}, obtain a non-metallic surface with two well separated surface bands (see Figure \ref{bands}); the occupied narrow one is mainly due to the C dangling bond , whereas the unoccupied wider one has a Si character as the projected densities of states reveals (see Figure \ref{Surface-Densities}). We, like Sabisch et al. \cite {Pollmann}, obtain a buckled non reconstructed surface; the C and Si atoms are displaced normal to the surface 0.01 \rm{\AA} outwards and 0.24 \rm{\AA} inwards respectively . The surface bonds are shorter (1.81 \rm{\AA}) than the bulk ones (1.91 \rm{\AA}) and the bond angles ($\sim {122^{\circ}}$) at the surface are larger than at the bulk ($ {109.47^{\circ}}$) making the surface chains more one-dimensional than the ideal structure ones.
We allowed in the calculation the possibility of different spin populations in the atoms. We did not  find a stable solution with non-zero magnetic moment at the surface atoms. It is worth to indicate that the densities of states at Figure \ref{Surface-Densities} suggest that introduction of holes in the system would shift the Fermi level to the sharp C-like peak giving rise eventually to a removal of states at the Fermi level by means of structural (Jahn-Teller type) and/or magnetic broken symmetries. However, introduction of electrons would shift the Fermi level to a much wider Si-like band, magnetic solutions being unlikely.
\begin{figure} 
\includegraphics[width=\columnwidth]
{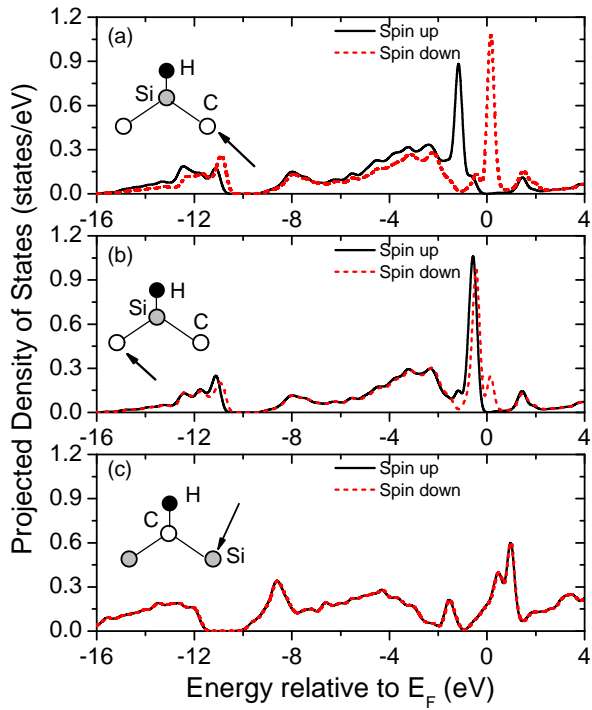}
\caption{(Color online) Projected spin resolved densities of states at the vicinity of an hydrogen atom chemisorbed at the SiC(110) surface. Solid (broken) lines represent spin up (down) states. Panels (a) and (b) stand for the densities of states projected on the C atom closest to the defect (H bonded to the surface Si atom). Panel (c) shows the densities of states projected on the Si atom closest to the defect (H bonded to the surface C atom) . The arrow at the insets indicate the atom where the density of states is projected. The origin of energies is taken at the Fermi Energy. A small imaginary contribution to the energy has been added for presentation purposes. }
\label{OntopSi}
\end{figure} 

We next consider different defects at the surface. We have considered a 2x2 surface unit cell with one defect in it. Nine (110) layers were included in the slab calculation. In most cases the results were checked by enlarging the supercell size. We first study a hydrogen atom saturating either a Si or a C surface atom.  In the case of hydrogen bonded to a Si atom we obtain, after full geometry relaxation, a magnetic solution such that the surface carbon atoms next to the defect have a magnetic moments of 0.705 $\mu_{B}$ and 0.058 $\mu_{B}$ respectively (see Table \ref{Table}).   This magnetic moment is very localized at one of the carbon atoms, the other carbon atoms in the unit cell have a magnetic moment of the order of 0.02 $\mu_{B}$ whereas the induced magnetic moments at the silicon atoms is also very small (see Table \ref{Table}). It should be indicated that the calculated total magnetic moment in the slab is 1 $\mu_{B}$. This magnetic solution is the stablest one with an energy gain of 0.142 eV per unit cell with respect to the diamagnetic one. The spin resolved projected densities of states are shown in Figure \ref{OntopSi}. The splitting of the carbon surface state peak due to the local magnetic moment is apparent.  It is interesting to realize that the magnetic moment symmetry of the two C atoms bonded to the defect is broken (see Table \ref{Table}) and, in addition, the symmetric geometry is also broken, including  a buckling of 0.22 \rm{\AA} of the C atoms. It is worth to note that while the magnetic solution is accompanied by this Jahn-Teller type distortion, a spin independent calculation does not give rise to any structural distortion. Our result clearly indicates an interesting lattice distortion local magnetic coupling. 
\begin{figure} 
\includegraphics[width=\columnwidth]
{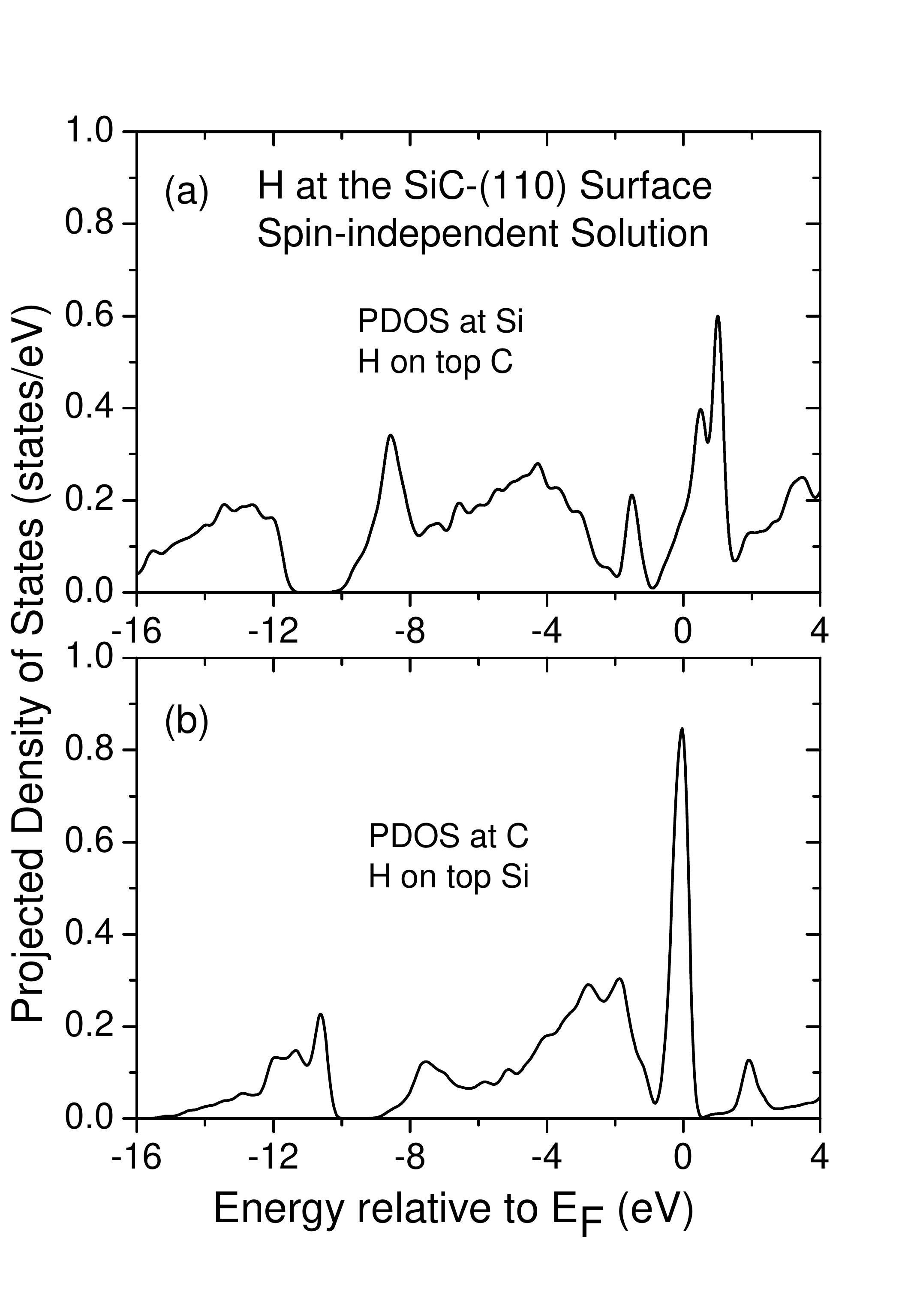}
\caption{Projected densities of states near adsorbed hydrogen in the spin independent solution. (a) Projected density of states at the Si surface atom next to the carbon atom bonded to the hydrogen impurity. 
(b)Projected density of states at the C surface atom next to the silicon atom bonded to the hydrogen impurity. In both cases the origin of energies is taken at the Fermi Energy} 
\label{Paramag}
\end{figure} 
The same calculations were performed in the case of H attached at the surface carbon atom. In this case, as anticipated, no magnetic stable solution was found obtaining a paramagnetic ground state. 
 \begin{table}[]
\caption {Charge (Q) and magnetic moment ($\mu$) at carbon atoms near different defects at the SiC(110) surface. The charges are the Mulliken populations. Q $\uparrow$,$\downarrow$ stand for the spin up and spin down electronic charge respectively. } 

		\begin{tabular}{cccccc}
\hline
\hline		
Atom & Q $\uparrow$ & Q $\downarrow$&  Qtot &$\mu$ ($\mu_{B}$)&Si-C (\rm{\AA}) \\

\hline
\hline 
H on top of Si\\			
\hline
Si bonded to H & 2.154 & 2.142&   4.296&0.012&- \\
\hline
C at Figure \ref {OntopSi}(a) & 2.179 & 1.474&3.653&0.705& 1.89\\
\hline
C at Figure \ref {OntopSi}(b) & 1.902&  1.844&3.746&0058&1.85\\
\hline 
\hline
F on top of Si\\	
\hline
Si bonded to F & 2.204 & 2.178&   4.382&0.026 \\
\hline
C next to Si & 2.144 & 1.491&   3.635&0.653 &1.87\\
\hline
C next to Si & 1.916 & 1.814 &  3.730&0.102&1.84\\
\hline
\hline Si Vacancy\\
\hline
C at Figure \ref{SiC-Vac-Si}(a)& 2.353 & 1.509&   3.862&0.843&-\\
\hline
C at Figure \ref{SiC-Vac-Si}(b)& 1.571 & 2.059&  3.630&-0.488&-\\
\hline
C at Figure \ref{SiC-Vac-Si}(a)& 2.353 & 1.509&   3.862&0.843&-\\
\hline
\hline Subst C\\
\hline
C next to extra C & 2.071 & 1.855& 3.926& 0.216&-\\
\hline
Extra C & 1.911 & 2.025&  3.936& -0.114&-\\
\hline	
C next to extra C & 2.071 & 1.855& 3.926& 0.216&-\\
\hline		 
\hline
\end{tabular}
\label{Table}	
\end{table}


In order to understand the origin of these localized moments when H is bonded to Si, and the absence of magnetic moments at the Si atoms when H is attached to C, we present in Figure \ref{Paramag} the results of the projected densities of a spin independent calculation. We first observe that in both cases the Fermi level lies at a band and therefore the system is metallic. However, the main difference between C and Si is that the density of states at the Fermi level in the case of H on top of Si is much higher than in the case of H on top of C. The C band is much narrower than the Si band due to the more localized valence orbitals in C than in Si, the corresponding covalent radii are 1.17 \rm{\AA} and 0.77 \rm{\AA} for Si and C respectively. This local character in the C case is also responsible for a larger electron-electron Hubbard like interaction U and therefore the fulfillment of the condition $U\times{N(E_{F})}> 1$, with $N(E_{F})$ the substrate density of states at the Fermi energy,
 to develop localized magnetic moments in a Stoner-like approach \cite{Anderson} .
\begin{figure} 
\includegraphics[width=\columnwidth]
{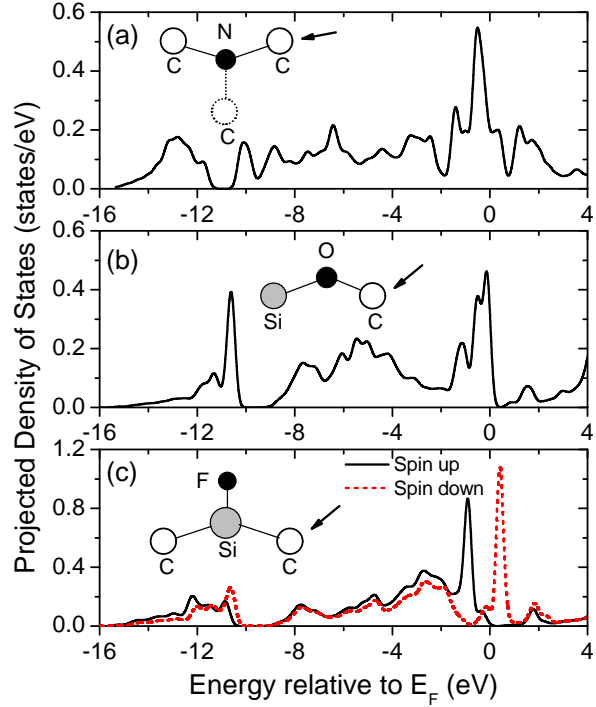}
\caption{(Color online) Projected densities of states near different impurities at the SiC(110) surface. The impurities are indicated by a black circle. The arrows indicate the C atoms where the density of states is projected. In the case of fluorine (panel (c)) both spin up (continuous black) and spin down (broken red) densities of states are shown. In the case of oxygen (panel (b)) and nitrogen (panel (a)) there is no distinction between spin up and spin down bands. The atomic configurations at the different defect geometries are sketched. The broken lines in panel (a) indicate the carbon atom under the topmost layer } 
\label{SiC-O-N-F}
\end{figure} 
 \begin{figure} 
\includegraphics[width=\columnwidth]
{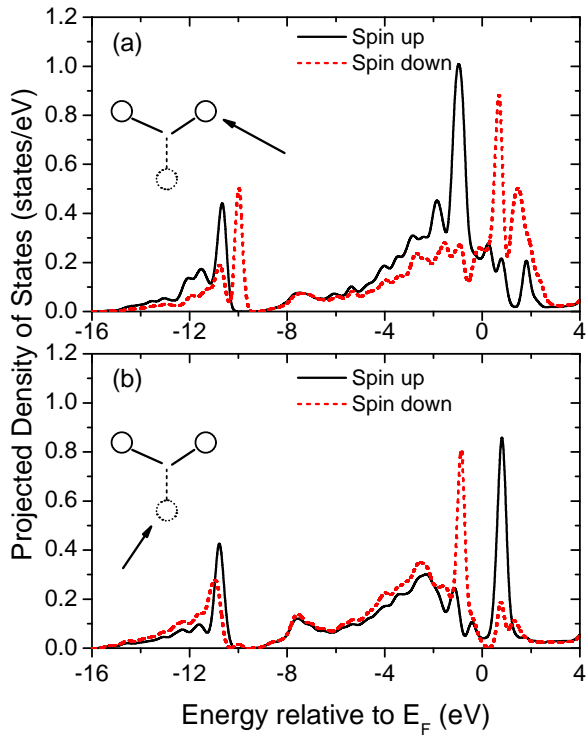}
\caption{(Color online) Projected densities of states near a Si vacancy at the SiC(110) surface. The surface C atoms are indicated by circles in the sketches. The broken lines in the sketches indicate the carbon atom under the topmost layer. The C atoms close the vacancy where the densities of states are projected are indicated by an arrow. (a) Densities of states at the outermost C atoms. (b) Densities of states at the C atom at the underneath layer. Solid black (broken red) lines stand for spin up (down) densities of states.} 
\label{SiC-Vac-Si}
\end{figure} 

Other possible surface defects were considered in detail:

i) A fluorine impurity. F behaves like H, it saturates the Si dangling bond and induces magnetism in the nearby C atoms (see  Table \ref{Table} and Figure \ref{SiC-O-N-F}(c)). The magnetic moments and atomic relaxations induced by F are very similar to those induced by H. Like in the case of H the total magnetic moment induced in the crystal is 1 $\mu_{B}$.

ii) An oxygen impurity. Oxygen can saturate the surface dangling bonds but energetically is much more favorable to break the Si-C surface bond rather than saturating the dangling bonds, like in the formation of SiO$_{2}$ at the Si surface \cite{Si-O-Si}. In this case, no magnetic solution was found. The corresponding projected density of states is drawn in Figure  \ref{SiC-O-N-F}(b).

iii) A substitutional nitrogen atom.  The most stable configuration of an N impurity is by substituting a Si surface atom and being bonded to three C atoms. The results of the calculation of this defect display no magnetic moment. The density of states at one of the surface carbon atoms bonded to N is shown in Figure  \ref{SiC-O-N-F}(a).
 
iv) A silicon vacancy. This configuration favors a magnetic solution at the C atoms (see Table \ref{Table}) such that the magnetic moment at the underneath carbon atom is in antiferromagnetic configuration with respect to the two surface C atoms. The total magnetic moment in the slab is zero. The projected densities of states at the carbon atoms close to the vacancy are shown in Figure  \ref{SiC-Vac-Si}.

 v) Substitutional carbon at the surface layer. In this case we also obtain a magnetic solution of the carbon atoms near the defect. The magnetic configuration of the carbon atoms is antiferromagnetic like in the case of the C-(110) surface \cite{C-(110)}. The total magnetic moment in the slab being also zero.

In conclusion, we have found that different defects at the SiC(110) surface can induce localized magnetic moments at the C atoms near the defect with various magnetic configurations. In the case of hydrogen or fluorine adsorbed at a Si surface atom magnetism is accompanied by a local lattice distortion. Magnetism takes place mainly at the carbon atoms due to the localized character of its atomic valence charge. We have found that the SiC(110) surface is an excellent candidate for magnetic configurations due to the quasi one dimensional behavior of the topmost Si-C chains. 

We thank J. M. Soler for helpful discussions about this work. 
Financial support of the Spanish Ministry of Science and Innovation through grants FIS2009-12712 and CSD2007-00050 is acknowledged.


\end{document}